\documentclass[twocolumn,showpacs,preprintnumbers,amsmath,amssymb,superscriptaddress]{revtex4}
\usepackage{graphicx}
\usepackage{dcolumn}
\usepackage{bm}

\begin{document}

%\maketitle
\title{Magnetic field induced pattern of coexisting condensates in the superconducting state of CeCoIn$_5$}%: the $\pi$-triplet scenario}

\author{Alexandros~Aperis}\author{Georgios~Varelogiannis}
\affiliation{Department of Physics, National Technical University of Athens,
GR-15780 Athens, Greece}
\author{Peter B.~Littlewood}
\affiliation{Cavendish Laboratory, University of Cambridge, Cambridge CB3 0HE, United Kingdom}

\vskip 0.7cm
%\pacs{}

\maketitle
%\begin{abstract}
{\bf
In usual superconductors the carriers form Cooper pairs \cite{BCS}
that have
zero total momentum meaning that the superfluid density is homogeneous.
We know for decades that it is a priori
possible to observe at high fields
the Fulde-Ferrel-Larkin-Ovchinikov (FFLO) \cite{FF,LO} state in which
the superfluid density exhibits a field dependent modulation.
The search for such inhomogeneous condensates in real materials
is the subject of tremendous excitement especially after the discovery
of a promising high-field superconducting (HFSC) phase in CeCoIn$_5$
\cite{Bianchi1}.
However, subsequent neutron \cite{Kenzelmann}
and NMR \cite{Mitrovic,Young} experiments
contradicted the FFLO picture in CeCoIn$_5$ establishing a puzzling coupling
of the distinct HFSC state with a magnetic modulation \cite{Kenzelmann}.
Here we show that, a novel type of exotic state is observed at high fields in CeCoIn$_5$
in which a pattern of coexisting condensates manifests. The specific pattern includes the d-wave singlet SC
state, the staggered $\pi$-triplet SC state and
Spin Density Waves (SDW). %\cite{Aperis}.
Because of particle-hole asymmetry these three condensates may either appear
separately or all three together providing a new perspective on
antiferromagnetic superconductors that may include high-$T_c$ cuprates and
pniktides.
The field induced transition to the pattern state in CeCoIn$_5$ is
only a paradigm of a generic
mechanism that prevents the elimination of the dominating condensate by the field
via the formation of a pattern of condensates. This
opens new avenues for the search of exotic states
not only in electronic systems, but also in
other systems where inhomogeneous
condensates are actively discussed like
trapped atomic fermi gases \cite{Giorgini}
and dense quark matter \cite{Casalbuoni}.}
%\end{abstract}

%\maketitle

Among the CeMIn$_5$ class of heavy-fermion superconductors
(M=Ir,Rh,Co), CeCoIn$_5$ exhibits the highest superconducting T$_c$
at ambient pressure (2.3K)\cite{Petrovic1}.
%Its 2D Fermi surface
%consists of quasi-cylindrical sheets as revealed by dHvA
%measurements\cite{Settai,Hall}. Moreover, CeCoIn$_5$ lies close to a
%magnetic quantum critical point exhibiting strong AF spin fluctuations \cite{Stock}.
The zero field SC gap symmetry is
established to be of d$_{x^2-y^2}$-wave type\cite{Izawa}.
A distinct HFSC state is observed initially thought to be
a realization of the FFLO state. Recent neutron diffraction data
%Kenzelmann {\it
%et al} \cite{Kenzelmann} have achieved neutron diffraction measurements
in the HFSC state of CeCoIn$_5$ show clearly that
%with the field along the
%$[1\bar{1}0]$ direction,
an almost commensurate SDW at {\bf
Q}=(q,q,0.5) develops at the onset of the HFSC region and disappears
at the same upper critical field with SC in a first order transition. Moreover, the
modulation wavevector \textit{is not coupled to
the magnitude of the external magnetic field} ruling out the FFLO mechanism
that produces superfluid density modulations that scale with the field.
The neutron results agree with previous NMR results
\cite{Mitrovic,Young} that reported SDW ordering
in the HFSC state and find no signature of normal state regions
required by the FFLO picture. Finally, in conflict with the FFLO expectations,
the critical
fields for entering the HFSC phase grow with temperature
and this behavior becomes more pronounced as we apply
pressure \cite{Miclea}.

The above facts
signal the
{\it field induced coexistence of
staggered $\pi-triplet$ SC with d-wave singlet SC and SDW}.
The staggered $\pi-triplet$ SC
component has some similarity with the FFLO state in the sense that
the pairs have a finite momentum and therefore the superfluid density is modulated
as well. However,
it is fundamentally different on basic points. Firstly, it is a {\it spin triplet}
SC state meaning that the paired quasiparticles have not antiparallel spin
as in usual Cooper pairs. Secondly, the wavevector
$Q$ of the superfluid modulation is driven by the
nesting properties of the dispersion and is common with that of the SDW modulation while
in the FFLO picture the wavevector is driven by the field.
Finally, our staggered triplet SC state coexists with the d-wave
singlet SC state and the SDW state, whereas the
FFLO state is in fact a coexistence between singlet SC and normal state regions on
different portions of the Fermi surface.

%The coexistence of singlet and triplet SC condensates would be a priori excluded
%because of the presence of inversion symmetry.
%However, our modulated superconducting triplet state has very unusual characteristics.
%Because {\bf Q} is {\it commensurate}, our d-wave singlet and
%staggered $\pi$-triplet SC states share {\it the same nodal structure} and
%therefore, their coexistence {\it is not contradicting the inversion symmetry}.
%Note that in the NMR experiments on CeCoIn$_5$ is identified a triplet
%component in the HFSC state that could not be understood because of the absence of
%a field induced change in the nodal structure \cite{Mitrovic}.

The possibility of a dynamically
generated $\pi$-triplet order parameter when
singlet SC coexists with SDW has already been considered
\cite{Aperis,Psaltakis,Murakami,Zhang,Kyung}.
%The
%concept of a $\pi$-operator has also been introduced in order to
%explain the resonance peak in the cuprates in terms of a collective
%excitation in the particle-particle channel of the Hubbard
%model\cite{Zhang}.
It has been shown \cite{Aperis, Tsonis} that
particle-hole asymmetry is enough to induce the third order
parameter {\it whenever the other two are present} and in this respect
the three condensates form a pattern \cite{Tsonis}.
Clearly, our findings may be relevant for any antiferromagnetic
superconductor and this includes high-$T_c$ cuprates and iron pnictides.
Moreover, the
staggered $\pi$-triplet SC state may also manifest in the ferromagnetic superconductor
UGe$_2$ \cite{Georgiou}.

\begin{figure}
\includegraphics[width=0.75\linewidth]{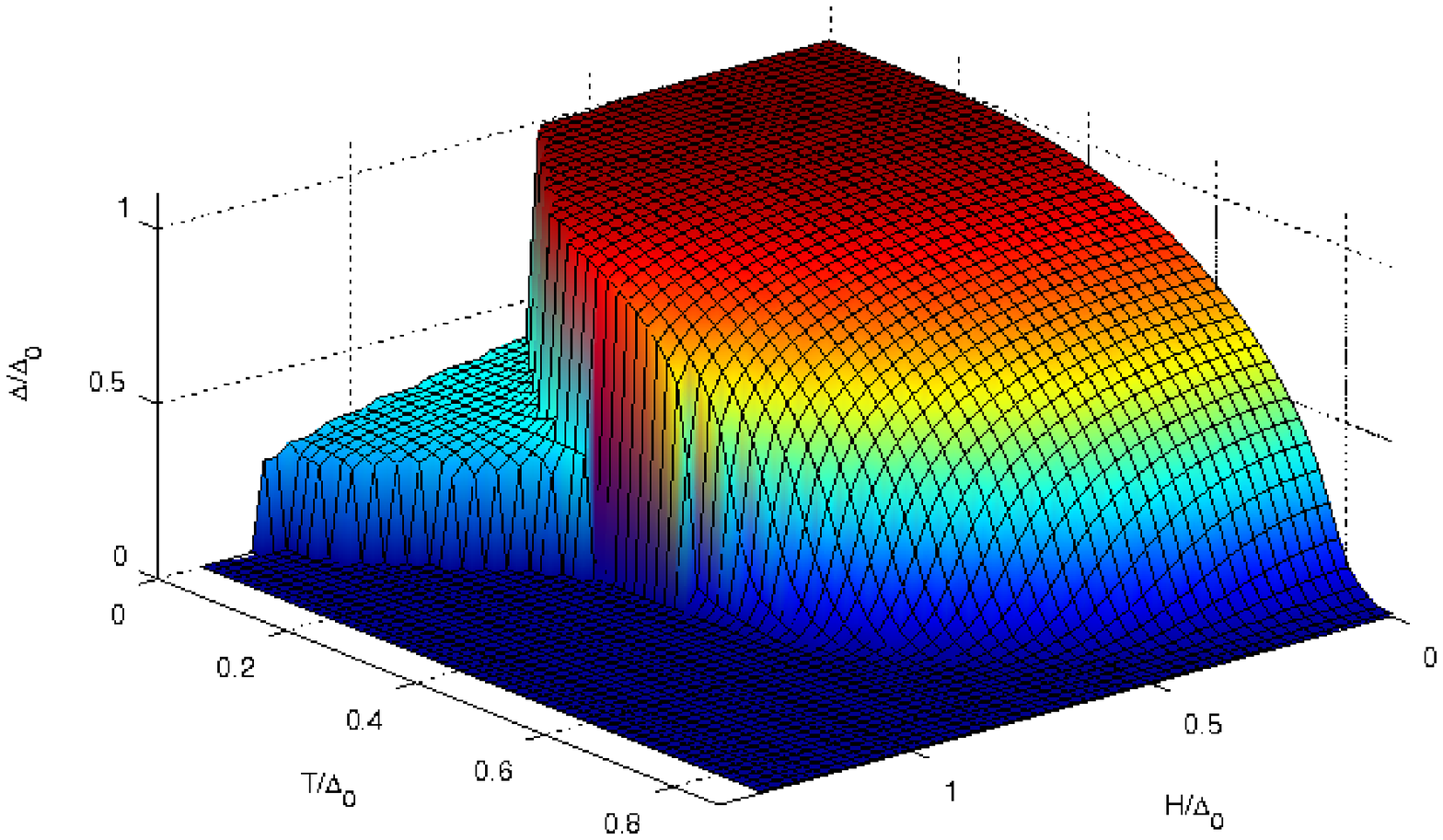}
\includegraphics[width=0.75\linewidth]{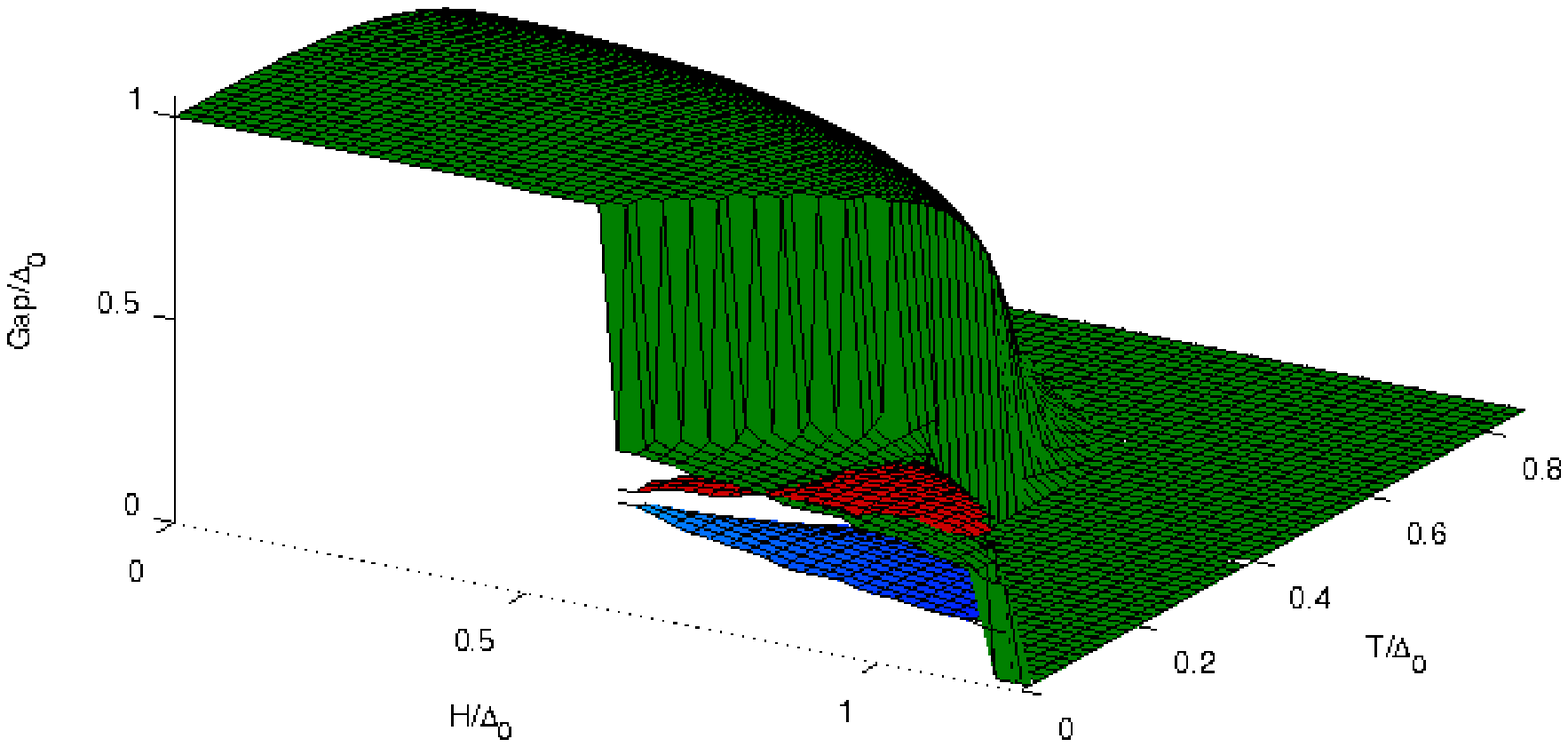}
\caption{\label{delta} In the upper panel is shown the field and temperature dependence of
the d-wave singlet SC gap ($\Delta$) when $t'=0.35t$ normalized to its value at zero temperature and field ($\Delta_0$). At low-T, a step-like regime bounded by 1st order transitions with the field takes place. In the lower panel we plot
the same as in the upper panel but for all three order parameters and a different orientation.
Here the d-wave singlet gap $\Delta$ is green, the staggered triplet SC order parameter $\Pi$
is red and the SDW order parameter M is blue.
In the low-T high field state {\it all three order parameters coexist}}.
%These results led to the H-T phase diagram in the main panel of Fig.~\ref{phd}}.
\end{figure}

Our analysis is absolutely generic, free of any model assumption.
We consider the simplest hamiltonian mean field scheme
that
includes the
relevant order parameters and the Zeeman field. The orbital effect of the field
is irrelevant for the phenomena that we report, and like for the FFLO states, results concern Pauli limited superconductors where
orbital effects are not dominating.
Our only requirement is to treat
the involved order parameters on the same footing. This is obtained using an eight component
spinor that leads to an $8\times 8$ matrix Green's function formalism
treated elsewhere \cite{Aperis}.
The poles of the Green's function are shown to be:
\begin{eqnarray}\label{qpoles}\nonumber
&&E_{\pm\pm}({{\bf k}})=H\pm\biggl[M^2+\gamma_{{\bf k}}^2+\delta_{{\bf k}}^2+\Delta_{{\bf k}}^2+\Pi_{{\bf k}}^2\pm\\
&&2\sqrt{\left(M^2+\gamma_{{\bf k}}^2\right)\delta_{{\bf k}}^2+2\delta_{{\bf k}}
M\Delta_{{\bf k}}\Pi_{{\bf k}}+\left(\gamma_{{\bf k}}^2+\Delta_{{\bf k}}^2\right)\Pi_{{\bf k}}^2}\biggl]^{1/2}
\end{eqnarray}
The letters $M,\Delta_{\bf k}$ and $\Pi_{{\bf k}}$ refer to the order parameters for the Spin Density Wave, d-wave singlet SC and $\pi$-triplet SC, respectively. The $\pi$-triplet order parameter corresponds to Cooper pairs having a finite total momentum equal to the nesting wavevector ${\bf Q}$. It has the attributes: $\Pi_{\bf k}=-\Pi_{\bf k+Q}=\Pi^*_{\bf k}=\Pi_{-\bf k}$ and because ${\bf Q}$ is commensurate this triplet SC order parameter is \textit{even in Parity}.
Indeed, our d-wave singlet and
staggered $\pi$-triplet SC states share {\it the same nodal structure} and
therefore, their coexistence {\it is not contradicting the inversion symmetry}.
The kinetic energy is decomposed into a sum of periodic $\delta_{\bf k+Q}=\delta_{\bf k}$ and antiperiodic $\gamma_{\bf k+Q}=-\gamma_{\bf k}$ terms on ${\bf Q}$ translations which is also the
commensurate nesting wavevector for the SDW. When $\delta_{\bf k}=0$ there is perfect
nesting and particle-hole symmetry. Finally H is the Zeeman field where
we have set $\mu_B=1$.

The resulting system of coupled self-consistent gap equations
is rather cumbersome exhibiting a peculiar
general structure:
\begin{eqnarray}\label{system}
M&=&\sum_n \sum_{{\bf k'}}V^{SDW}
\Bigl\{M' \bigl\{...\bigr\}+\delta_{\bf k'} \Delta_{\bf k'}
\Pi_{\bf k'}\bigl\{...\bigr\}\Bigr\} \nonumber \\ \Delta_{\bf k}
&=&\sum_n \sum_{{\bf k'}}V_{\bf k k'}^{dSC} \Bigl\{\Delta_{{\bf k'}}
\bigl\{...\bigr\}+\delta_{\bf k'} M'\Pi_{\bf k'}
\bigl\{...\bigr\}\Bigr\} \nonumber \\ \Pi_{\bf k} &=&\sum_n
\sum_{{\bf k'}}V_{\bf k k'}^{\pi tr} \Bigl\{\Pi_{{\bf k'}}
\bigl\{...\bigr\}+\delta_{\bf k'} M' \Delta_{\bf
k'}\bigl\{...\bigr\}\Bigr\} \end{eqnarray}where the $n$ sum is on the Matsubara frequencies and $V^{SDW},V^{dSC}_{\bf kk'},V^{\pi tr}_{\bf kk'}$ are the effective pairing potentials in the respective symmetry channels.
If the dispersion is particle-hole
asymmetric where $\delta_{\bf k}\neq 0$,
in none of the above equations we have zero as a self-consistent solution if the other two
order parameters are non-zero and naturally the potentials are non-zero as well.
In such a system we may either have one or all three order parameters simultaneously present. This is a generic result of relevance for all antiferromagnetic superconductors.
The three order parameters, constitute
a pattern of condensates.
Note that there are numerous other patterns
of different order parameters that behave similarly \cite{Tsonis}.
For example, SDW, CDW and FM form a similar pattern suggested as a possible
explanation for the CMR phenomenon in manganites \cite{varelogi2000}.

\begin{figure}
\includegraphics[width=0.75\linewidth]{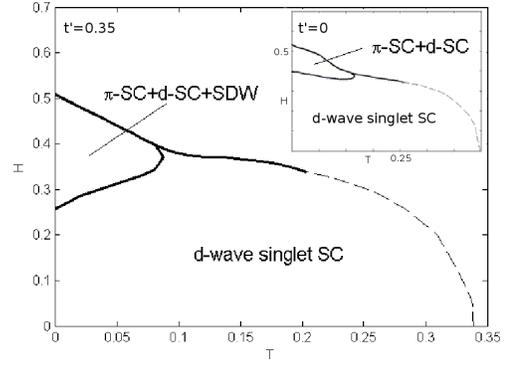}
\caption{\label{phd} Field-Temperature phase diagram for a particle-hole asymmetric system ($t'=0.35t$)
corresponding to Fig.~\ref{delta}. Solid lines mark 1st order and dashed lines 2nd order phase transitions.
%For low fields only d-wave singlet superconductivity is present.
At a sufficiently high field and low temperature, the system enters into a state in which all three order parameters coexist. The induced SDW lies only within the boundaries of the triplet $\bf{Q}$-modulated field induced superconductivity. Some similarities with a FFLO phase diagram are evident but here the critical field
for entering the multiphase state is enhanced with temperature. In the inset is shown the
corresponding phase diagram in the particle-hole symmetric case ($t'=0$). Note that in the particle-hole symmetric case there is no SDW order induced and the critical field of the step-like transition is rather temperature independent or even reduced with temperature.
The phase diagram in the main panel corresponds to the situation in CeCoIn$_5$ \cite{Kenzelmann}
whereas
that in the inset exhibits remarkable similarities with findings in organic
superconductors \cite{Singleton}}
\end{figure}

The system of self-consistent gap equations is solved numerically on a
square lattice with $\gamma_{\bf k}=-t\left(\cos{k_x}+\cos{k_y}\right)$ and $\delta_{\bf k}=-t'\cos{k_x}\cos{k_y}$ where  ${\bf Q}=(\pi,\pi)$. For convenience, separable pairing potentials of the form $V_{\bf kk'}=Vf_{\bf k}f_{\bf k'}$ have been used, where in our case $f_{\bf k}$ are d-wave form factors, $f_{\bf k}=\cos{k_x}-\cos{k_y}$. A thorough exploration of the combinations of potentials has been done numerically and we have found that there is a fairly big parameter space where the system is a d-wave singlet superconductor in the ground state. At a sufficiently high field, remarkable transitions to a mixed phase of all three order parameters are triggered. As a paradigm, we present in Fig.~\ref{delta}
 results obtained with $V^{SDW}=V^{dSC}=V^{\pi tr}=1.5t$
 and particle-hole asymmetry $t'=0.35$.
 %At low fields, $\Delta$ exhibits the usual BCS-like dependence with temperature via a 2nd order phase transition.
%For a sufficiently high field, the temperature transition becomes first order, which is indicated by the sudden drop of the gap at the critical temperature.
At low temperatures we observe two successive first order transitions with the field.
The first of them is from a high value of $\Delta$ to a smaller value of $\Delta$. In this step-like regime
the staggered triplet $\Pi$ gap and the SDW M gap appear simultaneously. At a higher field we have the second first order transition at which
all three order parameters are eliminated simultaneously.

\begin{figure}
\includegraphics[width=0.75\linewidth]{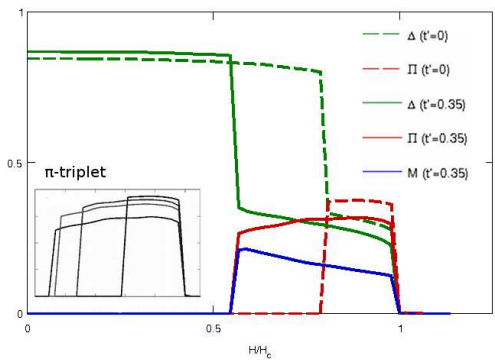}
\caption{\label{t2gap} Order parameter gap amplitudes as a function of the external field
in the particle-hole symmetric case $t'=0$ (dashed
lines) and in a particle-hole asymmetric case $t'=0.35$ (full lines).
With green, red and blue color we discern the d-wave singlet SC, the $\pi$-triplet SC and the SDW gap amplitude respectively. Notice how the coexistent regime is largely expanded with $t'$. To illustrate
this effect we plot in the inset the $\pi$-triplet gap as a function of field for $t'=(0,0.15,0.25,0.35)$. }
\end{figure}

\begin{figure}
\includegraphics[width=0.75\linewidth]{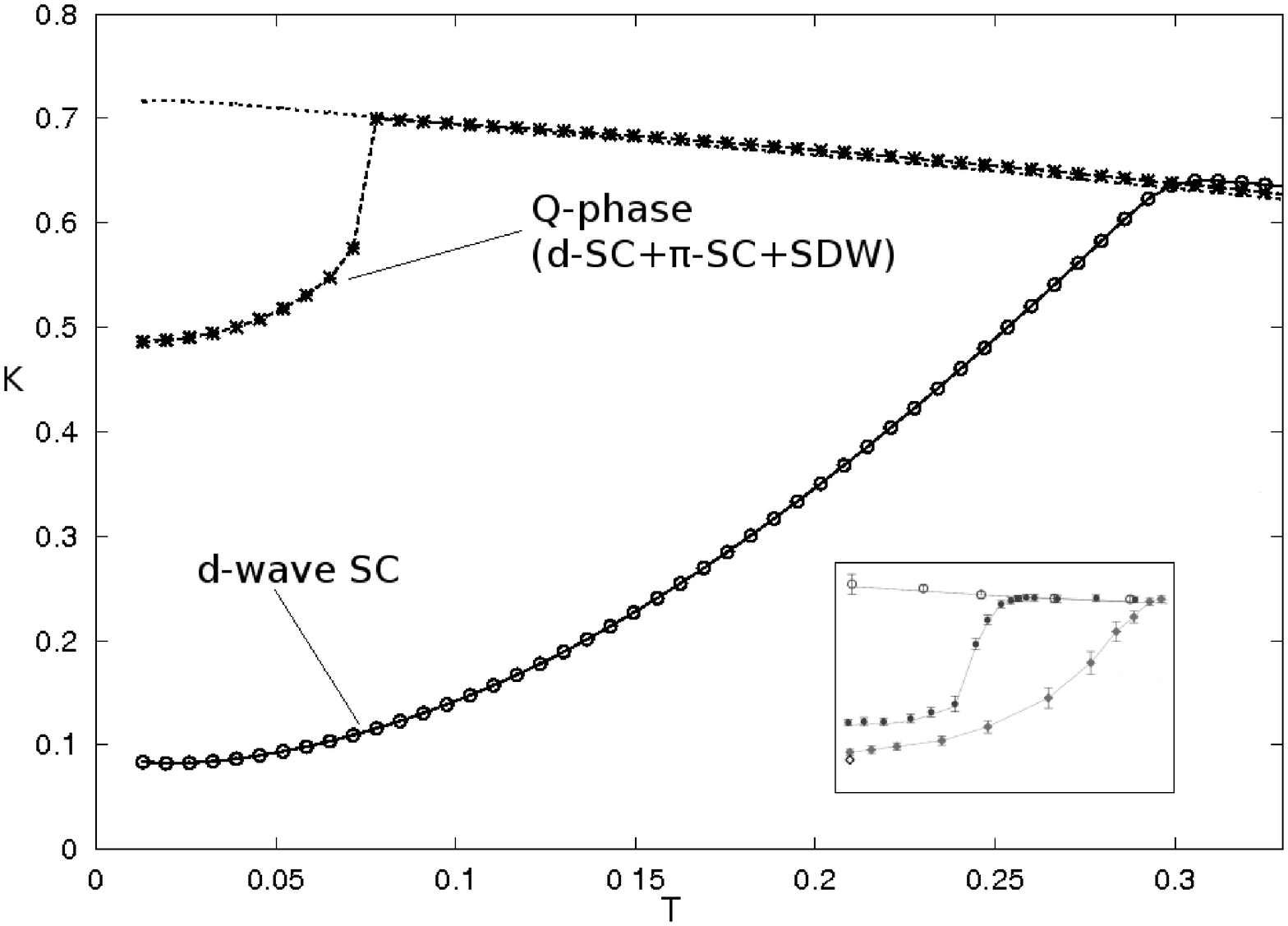}
\includegraphics[width=0.77\linewidth]{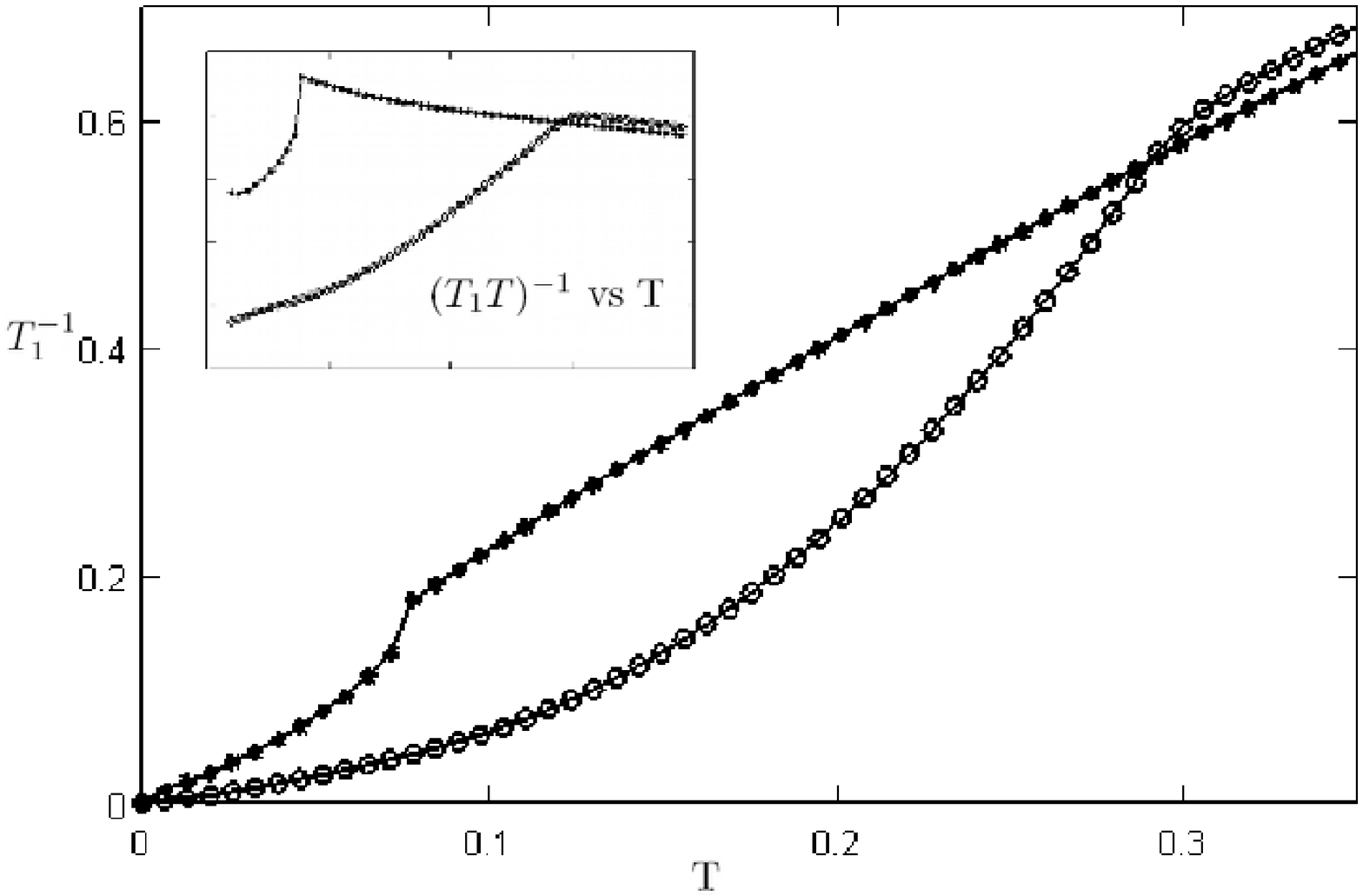}
\caption{\label{knight} In the upper panel is reported the Knight shift K as a function of temperature for three values of the Zeeman field corresponding to different regimes in our phase diagram. For a field corresponding to the modulated ${\bf Q}$-phase
at low temperatures, we obtain a temperature behavior almost identical to that
observed in NMR experiments \cite{Mitrovic} depicted schematically in the inset
of the upper panel.
At the critical temperature we observe a step in K associated with the first order transition from normal to the ${\bf Q}$-phase. For even lower temperatures the susceptibility is almost constant. In this state, the system is fully gapped in contrast with the FFLO state. The zero temperature value of the knight shift is higher in the the high field ${\bf Q}$-phase than in the d-wave singlet SC state (at lower fields)
signaling the presence of the staggered triplet superconducting component in the ${\bf Q}$-phase.
%Note the remarkable agreement of the
%knight shift calculations with the experimental results \cite{Mitrovic} depicted schematically in the inset
%of the upper panel.
In the lower panel is the corresponding inverse
NMR relaxation rate $(T_1T)^{-1}$ vs temperature for the same values of the external field
that lead to the d-wave singlet and the $\pi$-triplet SC transitions as in the upper panel. In the inset is
$1/T_1$ vs T .
}
\end{figure}

The resulting H-T phase diagram is reported in the main panel of Fig.~\ref{phd}.
The HFSC region in which the full pattern of coexisting
condensates is present corresponds to the $\bf{Q}$-phase in CeCoIn$_5$ \cite{Kenzelmann}.
%We must note here that although field induced first order transitions into the $\bf{Q}$-phase have been reported for CeCoIn$_5$, the transition is claimed to be of second order with temperature. This qualitiative feature is not captured by the present model, which does not include incommensurability or retardation effects.
%The triplet SC $\Pi$ and the SDW M gaps are
%finite only within this area.
At the highest critical field
the three order parameters are simultaneously eliminated in a first order transition
as observed by neutrons \cite{Kenzelmann}.

Remarkably, in the absence of particle-hole asymmetry $t'=0$
(inset of Fig.~\ref{phd}) {\it the SDW order is absent} from the field-induced state. The SDW
order is induced only because of particle-hole asymmetry given the peculiar structure
of the coupled self-consistent gap equations discussed above. The mechanism that drives the  ${\bf Q}$-phase in CeCoIn$_5$ is, therefore, the following:
to survive at the highest possible magnetic fields the
d-wave singlet SC condensate, develops an even in Parity, staggered in ${\bf Q}$ and odd in translation triplet superconducting component with the same nodal structure (i.e. the $\pi$-triplet component) with which is coexists. Particle-hole asymmetry {\it forces the presence of the third order parameter, the SDW condensate}. The full pattern of condensates appears at high fields when the dominating condensate, the d-wave singlet, is "weakened" sufficiently by the Zeeman field. Therefore, the emergence of the
pattern of condensates
can be viewed as an escapeway from  the formation of a magnetic quantum critical point
associated with the elimination of the dominant condensate.

Comparing the phase diagrams in Fig.~\ref{phd}, we note that the coexistence "dome"
\textit{expands with $t'$} as shown explicitly in Fig.~\ref{t2gap}.
The poles of the Green's function provide a straightforward explanation for this behavior.
%At low fields, our system is a d-wave superconductor, with the usual excitation spectrum:
%\begin{eqnarray}\label{dpoles}
%E_{\pm\pm}({{\bf k}})&=&H\pm\sqrt{\left(\gamma_{\bf k}\pm\delta_{\bf k}\right)^2+\Delta^2_{\bf k}}
%\end{eqnarray}
When $E_{--}({{\bf k}})=0$ the superconductor reaches its Clogston limit \cite{Clogston} and the transition to the multicomponent state is triggered. For higher $\delta_{\bf k}$,
smaller critical fields $H_c$ are necessary for reaching $E_{--}({{\bf k}})=0$ as confirmed by our numerical calculations. This behavior has been observed by applying pressure in CeCoIn$_5$ \cite{Miclea} and we may naturally simulate the effect of pressure as creating more deviations from nesting and therefore more particle-hole asymmetry $\delta_{\bf k}$.

Tedious calculations led to self-consistent NMR Knight shift and relaxation rates
as a function of temperature and field. Characteristic results reported in Fig.\ref{knight}
are for the same parameters as in Fig.~\ref{delta} for three characteristic fields ($H=0.21$, $H=0.41$ and $H=0.51$).
%In the upper panel
%is the temperature dependence of the Knight shift K for three characteristic fields ($H=0.21$, $H=0.41$ and $H=0.51$) showing respectively a transition to the d-wave singlet SC state, a transition to the ${\bf Q}$-phase and no transition as we lower the temperature.
There is a remarkable agreement with the experimental
results \cite{Mitrovic,Young} depicted in the inset of the upper panel.
Note the signature of the triplet component and the
signature of the first order transition to the Q-phase
in the self-consistent calculation and in the experiment.
The coexistence of singlet and triplet components evident in the NMR data
would be apriori excluded by the inversion symmetry. Only because
${\bf Q}$ is commensurate the staggered $\pi$-triplet condensate can coexist with
the d-wave singlet condensate with which it shares the same nodal structure.

%In the transition to the ${\bf Q}$-phase, the presence of the triplet component and the first order character of the transition are manifest
%natural results of the formation of the pattern of condensates.
%The fact that there were no normal regions observed inside the ${\bf Q}$-phase \cite{Mitrovic} despite the fact that the SC order parameter is modulated in real space, is simply due to the presence of all three order parameters making the system fully gapped.
%\begin{figure}
%\includegraphics[width=0.9\linewidth]{nm00.eps}
%\caption{\label{nmrrel}NMR relaxation rate vs temperature for two values of the external field for the d-wave %singlet and the $\pi$-triplet SC regime in correspondence with Figure \ref{knight}. In the main figure we %have plotted the $(T_1T)^{-1}$ as a function of T and in the inset $1/T_1$ vs T.}
%\end{figure}

%Finally, it is worth noting that in our formalism the wavevector of the momentum modulation
%${\bf Q}$ is the same for the the staggered triplet SC state and for the SDW and
%is bound to be commensurate. This wavevector is imposed by the nesting properties of the Fermi surface.
%As in the neutron data \cite{Kenzelmann}, the wavevector is not scaling with the magnetic
%field.
%Moreover, the observed modulation is nearly commensurate as we expect.

%For the first time is obtained definite evidence for the microscopic
%coexistence of more than two different condensates in an electronic system.
%The staggered $\pi$-triplet superconducting state has never been identified
%before in a real system.

The field induced transition to the pattern state is obtained through an
absolutely generic mean field scheme. Therefore, field induced patterns have
a generic character and
should be viewed as
potential alternatives to the FFLO picture not only in superconductors.
Complexity in electronic matter is proven to be crucially important.
%but also in other systems where such inhomogeneous condensates are invoked.
%For example in charge density wave (CDW) systems at high magnetic fields
%similar field induced transitions
%to a CDW+SDW state \cite{AperisEPL} may occur \cite{McDonald}.
Our findings provide a new perspective in research areas as diverse as
the trapped Fermi gases \cite{Giorgini}
and situations of dense quark matter in QCD \cite{Casalbuoni}
where
inhomogeneous condensates are very actively investigated.
%Moreover, in simple charge density wave (CDW) systems there are
%already reports for such transitions
%that may correspond to a different pattern of condensates that involve CDW, SDW and ferromagnetism \cite{varelogi2000,EPLaperis}.
%In general,
%the so called "dome" states close to a magnetic quantum critical point may acquire a new
%sense and become the ideal ground for the search of exotic condensates.

We acknowledge discussions with Piers Coleman, Panagiotis Kotetes, Efthimios Liarokapis,
Gil Lonzarich, Peter Oppeneer, Ben Simons, Peter Thalmeier, Stefan Wirth and Hiroyuki Yamase. This work has been funded by the European Union through the STRP
NMP4-CT-2005-517039  CoMePhS project.

\end{document}